\documentclass{article}
\usepackage{amssymb}

\usepackage{amsmath}
\usepackage{mathrsfs}
\usepackage{epsfig}
\usepackage{epstopdf}

\newtheorem{theorem}{Theorem}

\newtheorem{condition}[theorem]{Condition}

\newtheorem{definition}[theorem]{Definition}

\newtheorem{lemma}[theorem]{Lemma}

\newtheorem{proposition}[theorem]{Proposition}

\begin{document}

\title{Zeno Dynamics for Open Quantum Systems}
\date{}
\author{John Gough\\
Department of Mathematics and Physics,\\
 Aberystwyth University, SY23 3BZ, Wales, UK}
\date{}
\maketitle

\begin{center}
Dedicated to the memory of Slava Belavkin, the father of quantum cybernetics.
\end{center}

\begin{abstract}
In this paper we formulate limit Zeno dynamics of general open systems as the adiabatic elimination of fast components. 
We are able to exploit previous work on adiabatic elimination of quantum stochastic models to give explicitly the conditions
under which open Zeno dynamics will exist. The open systems formulation is further developed as a framework for Zeno master
equations, and Zeno filtering (that is, quantum trajectories based on a limit Zeno dynamical model). We discuss several models from
the point of view of quantum control. For the case of linear quantum stochastic systems we present a condition for stability of the 
asymptotic Zeno dynamics.
\end{abstract}

\section{Introduction}
The quantum Zeno effect is the basic principle that repeated or continual external interventions on a quantum dynamical system may lead to a
constraining of the system to a sub-dynamics. The original formulation by
Misra and Sudarshan \cite{MS77} dealt with a system initially in an unstable system $|\psi _{0}\rangle $ which is repeatedly measured to see if it is in this
state (a von Neumann projective measurement with orthogonal projection $P_{\mathtt{z}}=|\psi _{0}\rangle \langle \psi _{0}|$). At each stage the survival
probability $p(t)=|\langle \psi _{0}|e^{-itH}\psi _{0}\rangle |^{2}$ is $1+O(t^{2})$ for small $t$, so for $N$ measurements at time separations $t/N$
the probability of measuring the same state on all $N$ attempts is $p(t/N)^{N}$ 
which converges to $1$ as $N\uparrow \infty $. The state is in principle ``frozen'' as $|\psi _{0}\rangle $, up to a phase.

More generally the projection $P_{\mathtt{z}}$ may be of rank greater than one, in which case we find that the state is constrained to remain the Zeno subspace
\begin{equation*}
\mathfrak{h}_{\mathrm{Zeno}}=P_{\mathtt{z}} \, \mathfrak{h}
\end{equation*}
where $\mathfrak{h}$ is the total Hilbert space of states. Typically, we still have a reduced evolution on the Zeno subspace governed by the Zeno
Hamiltonian $H_{\mathrm{Zeno}}=P_{\mathtt{z}}HP_{\mathtt{z}}$. (The unitary $e^{-itP_{\mathtt{z}}HP_{\mathtt{z}}}$ on $\mathfrak{h}_{\mathrm{Zeno}}$ sometimes being referred to as a ``non-abelian phase''). 
One can obtain results of the form
\begin{equation*}
\lim_{N\rightarrow \infty }\left[ P_{\mathtt{z}}e^{-iHt/N}\right] ^{N}=P_{\mathtt{z}}e^{-itP_{\mathtt{z}}HP_{\mathtt{z}}}
\end{equation*}

The intervention does not necessarily have to take the form of a
measurement, see the authoritative review by Facchi and Pascazio \cite{FP}. An alternative is to apply stroboscopic unitary kicks and
take the limit as the time between kicks vanishes. Another possibility is to
apply a strong perturbation $kV$ to obtain a result of the form 
\begin{equation*}
\lim_{k\rightarrow \infty }e^{+ikVt}e^{-i\left( H+kV\right) t}P_{\mathtt{z}}=P_{\mathtt{z}}e^{-itP_{\mathtt{z}}HP_{\mathtt{z}}}
\end{equation*}
where $P_{\mathtt{z}}$ is projection onto the kernel of $V$. This is a case of the
adiabatic theorem.

Despite the fact that quantum Zeno effect is typically described in terms of
open systems concepts such as quantum measurement, external interaction or
environmental coupling, it is rarely if ever formulated mathematically in
these terms. In practice, on never makes direct measurements on quantum
mechanical systems, but instead couples them to external quantum fields
which are then monitored continually. A complete model would include the
probing field \cite{Bel80}-\cite{Bel92a}. Similarly, the coupling of the environment requires an open
systems model.

\begin{figure}[htbp]
	\centering
		\includegraphics[width=1.00\textwidth]{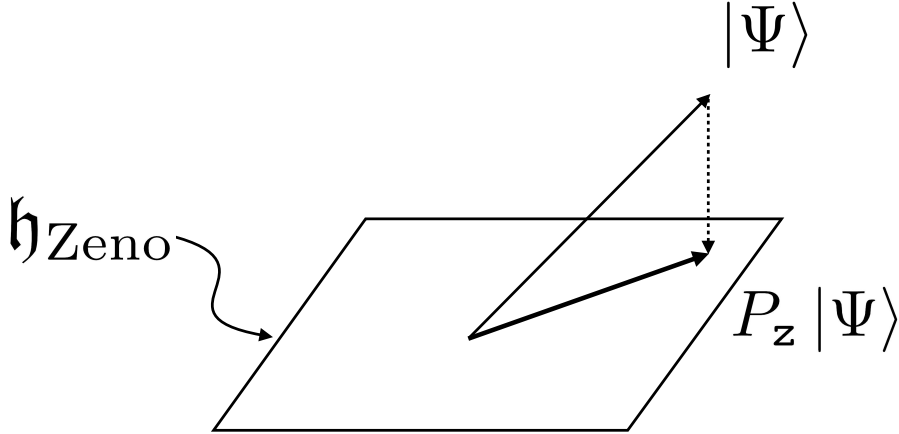}
	\caption{In a Zeno limit we find that a quantum state $\vert \Psi \rangle$ will have component perpendicular to a Zeno subspace $\mathfrak{h}_{\mathrm{Zeno}}$ which so strongly coupled to the environment that it effectively disappears in real time, leaving the system constrained to the Zeno subspace. The component may be adiabatically eliminated.}
	\label{fig:Zeno}
\end{figure}

In this paper we start in the general framework of an open quantum Markov
model for the system undergoing a quantum stochastic evolution in
conjunction with a Boson field environment \cite{HP,partha,Gardiner}. We shall equate the limit
procedure leading to a Zeno dynamics as adiabatic elimination. This choice
is a matter of convenience as we at once have access to powerful results due
to Luc Bouten and Andrew Silberfarb \cite{Bouten_Silberfarb}, and later with Ramon van Handel 
\cite{Bouten_Handel_Silberfarb}, on the
adiabatic elimination procedure in the framework of quantum stochastic
models. This is a significantly richer theory than for closed systems. In
addition we get clear conditions on when a candidate subspace will be a Zeno
subspace for a sequence of models. We can the readily develop the theory of
how we actually probe the Zeno sub-dynamics by for instance quantum
trajectories methods for the constrained dynamics \cite{Belavkin} - \cite{BHJ07}.

\subsection{Open Quantum Systems: The ``$S,L,H$'' Formalism}

In figure \ref{input-plant-output component} below, we sketch an open quantum Markov
 system as a black box with a Bosonic input driving field
and a Bosonic output. The quantum
mechanical system (plant) will have underlying Hilbert space $\mathfrak{h}$
while the input will be a continuous quantum field with Fock space $\mathfrak{F}$. The coupled model will have joint Hilbert space $\mathfrak{h}\otimes \mathfrak{F}$, 
which is also the space on which the output observables act.

We shall outline below the theory of quantum stochastic evolutions due to Hudson-Parthasarathy. 
This shows that the unitary dynamics of the combined system and its Bosonic environment are specified by a triple of coefficients 
\begin{equation}
\mathbf{G} \sim (S,L,H) .
\end{equation}

We shall in turns refer to $\mathbf{G}$ as the Hudson-Parthasarathy (HP) parameters, the generating coefficients, or just the $SLH$ for the model.

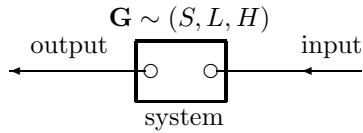
\begin{figure}[h]
\centering
\begin{picture}(120,55)
\setlength{\unitlength}{.04cm}%
\label{pic1}
\thicklines
\put(45,10){\line(0,1){20}}
\put(45,10){\line(1,0){30}}
\put(75,10){\line(0,1){20}}
\put(45,30){\line(1,0){30}}
\thinlines
\put(48,20){\vector(-1,0){45}}
\put(120,20){\vector(-1,0){20}}
\put(120,20){\line(-1,0){48}}
\put(50,20){\circle{4}}
\put(70,20){\circle{4}}
\put(100,26){input}
\put(36,35){$\mathbf{G} \sim (S,L,H)$}
\put(48,2){system}
\put(10,26){output}
\end{picture}
\caption{input-system-output component}
\label{input-plant-output component}
\end{figure}

The input-plant-output model can be summarized as 
\begin{eqnarray*}
\mathbf{plant~dynamics} &:&j_{t}\left( X\right) =U\left( t\right) ^{\ast
}\left( X\otimes I\right) U\left( t\right) ; \\
\mathbf{output~process} &:&B_{\mathrm{out},i}\left( t\right) =U\left(
t\right) ^{\ast }\left( I\otimes B_{i}\left( t\right) \right) U\left(
t\right) .
\end{eqnarray*}
where $X$ is an arbitrary plant observable and $B_{i}\left( t\right) $ is
the $i$th input field's annihilator process.

In the following we shall specify to the category of model where $U\left(
\cdot \right) $ is a unitary family of operators on $\mathfrak{h}\otimes 
\mathfrak{F} $, satisfying a differential equation of the form \cite{HP,partha} 
\begin{eqnarray}
dU\left( t\right) = \left\{ \sum_{ij}\left( S_{ij}-\delta _{ij}\right)
\otimes d\Lambda _{ij}\left( t\right) +\sum_{i}L_{i}\otimes dB_{i}^{\ast
}\left( t\right) \right.  \notag \\
\left. -\sum_{ij}L_{i}^{\ast }S_{ij}\otimes dB_{j}\left( t\right) +K\otimes
dt\right\} U\left( t\right) ,\quad U(0)=I,  \label{HP_QSDE}
\end{eqnarray}
Formally, we can introduce input process $b_{i}\left( t\right) $ for $i=1,\cdots ,n$ satisfying singular commutation relations 
$[ b_{ i}(t),b_{j}(t^{\prime })^{\ast }]=\delta _{ij}\delta (t-t^{\prime })$, so that the
processes appearing in are 
\begin{eqnarray*}
\Lambda _{ij}\left( t\right) &\triangleq &\int_{0}^{t}b_{i}\left( t^{\prime
}\right) ^{\ast }b_{ j}(t^{\prime })dt^{\prime }, \\
B_{i}\left( t\right) ^{\ast } &\triangleq &\int_{0}^{t}b_{i}\left( t^{\prime
}\right) ^{\ast }dt^{\prime },\quad B_{j}\left( t\right) \triangleq
\int_{0}^{t}b_{ j}(t^{\prime })dt^{\prime }.
\end{eqnarray*}
More exactly, the are rigorously defined as creation and annihilation field
operators on the Boson Fock space $\mathfrak{F}$ over $L_{\mathbb{C}%
^{n}}^{2}\left( \mathbb{R}\right) $. The increments in \ are understood to
be future pointing in the Ito sense. We have the following table of
non-vanishing products 
\begin{eqnarray*}
d\Lambda _{ij}d\Lambda _{kl} &=&\delta _{jk}d\Lambda _{il},\qquad d\Lambda
_{ij}dB_{k}^{\ast }=\delta _{jk}dB_{i}^{\ast } \\
dB_{i}d\Lambda _{kl} &=&\delta _{ik}dB_{l},\qquad dB_{i}dB_{k}^{\ast
}=\delta _{ij}dt.
\end{eqnarray*}

Necessary and sufficient conditions for unitarity \cite{HP,partha} are that we can collect the coefficients of (\ref{HP_QSDE}) to form a triple $(S,L,H)$, 
which we call the \emph{Hudson-Parthasarathy (HP) or $SLH$ parameters}, consisting of a unitary matrix $S$, a column vector $L$, and a self-adjoint operator $H$, 
\begin{equation*}
S=\left[ 
\begin{array}{ccc}
S_{11} & \cdots & S_{1n} \\ 
\vdots & \ddots & \vdots \\ 
S_{n1} & \cdots & S_{nn}
\end{array}
\right] , \quad L=\left[ 
\begin{array}{c}
L_{1} \\ 
\vdots \\ 
L_{n}
\end{array}
\right] ,
\end{equation*}
with $S_{ij},L_{i},H$ are all operators on $\mathfrak{h}$, and where 
\begin{equation*}
K\equiv -\frac{1}{2}\sum_{i}L_{i}^{\ast }L_{i}-iH.
\end{equation*}

We shall refer to $U(t)$ as the unitary determined by the parameters $\left(
S,L,H\right) $. In differential form, the input-plant-output model then
becomes \cite{HP,partha}

\bigskip

\noindent \textbf{plant dynamical (Heisenberg) equation:} 
\begin{eqnarray*}
dj_{t}\left( X\right) &=&j_{t}\left( \mathscr{L}X\right)
dt+\sum_{i}j_{t}\left( \mathscr{M}_{i}X\right) dB_{i}^{\ast }\left( t\right)
\\
&&+\sum_{i}j_{t}\left( \mathscr{N}_{i}X\right) dB_{i}\left( t\right)
+\sum_{j,k}j_{t}\left( \mathscr{S}_{jk}X\right) d\Lambda _{jk}\left(
t\right) ;
\end{eqnarray*}

\bigskip

\noindent \textbf{input-output relations:}

\begin{equation*}
dB_{\mathrm{out},i}\left( t\right) =j_{t}\left( S_{ik}\right) dB_{k}\left(
t\right) +j_{t}\left( L_{i}\right) dt.
\end{equation*}

Here 
\[
\mathscr{L}X=\frac{1}{2}\sum_{i}L_{i}^{\ast }\left[ X,L_{i}\right] +\frac{1}{2}\sum_{i}\left[ L_{i}^{\ast },X\right] L_{i}-i\left[ X,H\right] 
\]
is the Lindbladian, $\mathscr{M}_i X = S_{ji}^\ast [X,L_j ], \mathscr{N}_i X= [L_k^\ast , X ] S_{ki}$ and $\mathscr{S}_{ik} X = S_{ji}^\ast XS_{jk} $
$-\delta_{ik} X$.

\section{Zenofiability}

The essential idea is that there exists a subspace $\mathfrak{h}_{\mathrm{Zeno}}$ of $\mathfrak{h}$ to which the system may end up restricted (though
entangled with the environment) due to the continuous open system dynamics.
We decompose $\mathfrak{h}$ as 
\begin{equation}
\mathfrak{h}=\mathfrak{h}_{\mathrm{Zeno}}\oplus \mathfrak{h}_{\mathrm{Fast}}
\label{fs_decomp}
\end{equation}
where the subspace $\mathfrak{h}_{\mathrm{Fast}}=\mathfrak{h}\ominus \mathfrak{h}_{\mathrm{Zeno}}$ contains the vector states that are most
strongly coupled to the environment so that any drift out of $\mathfrak{h}_{\mathrm{Zeno}}$ will decay to zero almost instantaneously.

To make these concepts mathematically rigorous, we introduce a strength parameter $k>0$ for the coupling and make the generating operators $k
$-dependent according to $\mathbf{G} (k)$ with $SLH$ operators of the form
\begin{eqnarray}
S(k) &=&S\text{, \ \ (independent of }k) ;  \notag \\
L\left( k\right)  &=&kL^{\left( 1\right) }+L^{\left( 0\right) };  \notag \\
H\left( k\right)  &=&k^{2}H^{\left( 2\right) }+kH^{\left( 1\right)
}+H^{\left( 0\right) },  \label{eq:scaling1}
\end{eqnarray}
leading to the open unitary dynamics $U(t,k)$.

The singular limit $k\uparrow \infty $ should then separate $\mathfrak{h}_{\mathrm{Fast}}$ out as the fast subspace which may be eliminated to yield a
dynamics on the open system over $\mathfrak{h}_{\mathrm{Zeno}}$. At this stage, the Zeno limit can be reformulated explicitly as an adiabatic
elimination problem with the Zeno subspace $\mathfrak{h}_{\mathrm{Zeno}}$ being the slow space. At this stage we can invoke the results of Bouten, van
Handel and Silberfarb \cite{GvH},\cite{Bouten_Silberfarb},\cite{Bouten_Handel_Silberfarb} on adiabatic elimination for quantum stochastic systems.

For a given operator $X$ on $\mathfrak{h}$, we write 
\begin{equation*}
X=\left[ 
\begin{array}{cc}
X_{\mathtt{zz}} & X_{\mathtt{zf}} \\ 
X_{\mathtt{fz}} & X_{\mathtt{ff}}
\end{array}
\right] .
\end{equation*}
More generally we use this notation when $X$ is an array of operators on $\mathfrak{h}$. The projections onto $\mathfrak{h}_{\mathrm{Zeno}}$ and 
$\mathfrak{h}_{\mathrm{Fast}}$ are denoted respectively by 
\[P_{\mathtt{z}}\equiv \left[ 
\begin{array}{cc}
1 & 0 \\ 
0 & 0
\end{array}
\right] , \quad
P_{\mathtt{f}}\equiv \left[ 
\begin{array}{cc}
0 & 0 \\ 
0 & 1
\end{array}
\right] .
\]

\subsection{Conditions}

\bigskip 

\begin{condition}[Scaling]
We assume that the $SLH$ operators in (\ref{eq:scaling1})\ satisfy
\begin{equation}
L^{\left( 1\right) }P_{\mathtt{z}}=0,\quad H^{\left( 1\right) }P_{\mathtt{z}}=0,\quad P_{\mathtt{z}}H^{\left( 1\right) }=0,\quad P_{\mathtt{z}}H^{\left(
2\right) }P_{\mathtt{z}}=0.
\end{equation}
\end{condition}

Equivalently stated we require the following forms with respect to the decomposition:
\begin{eqnarray*}
L^{\left( 1\right) } &\equiv &\left[ 
\begin{array}{ll}
0 & L_{\mathtt{zf}}^{\left( 1\right) } \\ 
0 & L_{\mathtt{ff}}^{\left( 1\right) }
\end{array}
\right] , \\
H &\equiv &\left[ 
\begin{array}{ll}
H_{\mathtt{zz}}^{\left( 0\right) } & H_{\mathtt{zf}}^{\left( 0\right) }+kH_{\mathtt{zf}}^{\left( 1\right) } \\ 
H_{\mathtt{fz}}^{\left( 0\right) }+kH_{\mathtt{fz}}^{\left( 1\right) } & H_{\mathtt{ff}}^{\left( 0\right) }+kH_{\mathtt{ff}}^{\left( 1\right) }+k^{2}H_{%
\mathtt{ff}}^{\left( 2\right) }
\end{array}
\right] .
\end{eqnarray*}
In what follows, it is instructive to write
\begin{eqnarray*}
S &=&\left[ 
\begin{array}{cc}
S_{\mathtt{zz}} & S_{\mathtt{zf}} \\ 
S_{\mathtt{fz}} & S_{\mathtt{ff}}
\end{array}
\right] , \\
L(k) &=&k\left[ 
\begin{array}{cc}
0 & L_{\mathtt{zf}}^{(1)} \\ 
0 & L_{\mathtt{ff}}^{(1)}
\end{array}
\right] +\left[ 
\begin{array}{cc}
L^{\left( 0\right) }_{\mathtt{zz}} & \ast  \\ 
L^{(0)}_{\mathtt{fz}} & \ast 
\end{array}
\right] , \\
H(k) &=&k^{2}\left[ 
\begin{array}{cc}
0 & 0 \\ 
0 & H_{\mathtt{ff}}^{(2)}
\end{array}
\right] +k\left[ 
\begin{array}{cc}
0 & H_{\mathtt{zf}}^{(1)} \\ 
H^{\left( 1\right) }_{\mathtt{fz}} & \ast 
\end{array}
\right] +\left[ 
\begin{array}{cc}
H^{(0)}_{\mathtt{zz}} & \ast  \\ 
\ast  & \ast 
\end{array}
\right] 
\end{eqnarray*}
where the terms denote as $\ast $ will not contribute to the $k\uparrow
\infty $ limit. We also note that 
\begin{equation*}
K\left( k\right) =-\frac{1}{2}L(k)^{\ast }L(k)-iH(k)\equiv k^{2}A+kM+R,
\end{equation*}
where we will have
\begin{equation*}
A=\left[ 
\begin{array}{ll}
0 & 0 \\ 
0 & A_{\mathtt{ff}}
\end{array}
\right] ,M=\left[ 
\begin{array}{ll}
0 & M_{\mathtt{zf}} \\ 
M_{\mathtt{fz}} & \ast 
\end{array}
\right] ,R=\left[ 
\begin{array}{ll}
R_{\mathtt{zz}} & \ast  \\ 
\ast  & \ast 
\end{array}
\right] 
\end{equation*}
with 
\begin{eqnarray*}
A_{\mathtt{ff}} &=&-\frac{1}{2}L_{\mathtt{af}}^{\left( 1\right) \ast }L_{\mathtt{af}}^{\left( 1\right) }-iH_{\mathtt{ff}}^{\left( 2\right) } \\
M_{\mathtt{zf}} &=&-\frac{1}{2}L_{\mathtt{cz}}^{\left( 0\right) \ast }L_{\mathtt{cf}}^{\left( 1\right) }-iH_{\mathtt{zf}}^{\left( 1\right) } \\
M_{\mathtt{fz}} &=&-\frac{1}{2}L_{\mathtt{cf}}^{\left( 1\right) \ast }L_{\mathtt{cz}}^{\left( 0\right) }-iH_{\mathtt{fz}}^{\left( 1\right) } \\
R_{\mathtt{zz}} &=&-\frac{1}{2}L_{\mathtt{cf}}^{\left( 0\right) \ast }L_{\mathtt{cz}}^{\left( 0\right) }-iH_{\mathtt{zz}}^{\left( 0\right) }
\end{eqnarray*}
and we introduce a summation convention where repeated indices imply a summation over the values $\mathtt{z}$ and $\mathtt{f}$.

\begin{condition}[Kernel]
We require  $\mathfrak{h}_{\mathrm{\mathtt{z}}}$ to be the kernel space of $A$.
\end{condition}

Equivalently, we require that the operator $A_{\mathtt{ff}}$ is invertible
on $\mathfrak{h}_{\mathrm{\mathtt{z}}}$. Making this assumption we then
introduce the operators
\begin{eqnarray}
\hat{S}_{\mathtt{ab}} &\triangleq &\left( \delta _{\mathtt{ac}}+L_{\mathtt{af}}^{\left( 1\right) }\frac{1}{A_{\mathtt{ff}}}L_{\mathtt{cf}}^{\left(
1\right) \ast }\right) S_{\mathtt{cb}}, \\
\hat{L}_{\mathtt{a}} &\triangleq &L_{\mathtt{az}}^{\left( 0\right) }-L_{\mathtt{af}}^{\left( 1\right) }\frac{1}{A_{\mathtt{ff}}}M_{\mathtt{fz}}, \\
\hat{H} &\triangleq &H_{\mathtt{zz}}^{(0)}+\mathrm{Im}\left\{
M_{\mathtt{zf}}\frac{1}{A_{\mathtt{ff}}}M_{\mathtt{fz}}\right\} .
\end{eqnarray}

With these definitions we can state the final condition.

\begin{condition}[Decoupling]
The operators $\hat{S}_{\mathtt{zf}},\hat{S}_{\mathtt{fz}}$ and $\hat{L}_{\mathtt{f}}$\ vanish.
\end{condition}

\subsection{Asymptotic Open Zeno Dynamics}

We now recall the adiabatic elimination result of Bouten and Silberfarb
which establishes convergence to a limit unitary $\hat{U}(t)$ with reduced
system space $\mathfrak{h}_{\mathrm{Zeno}}$ and $SLH$ operators 
\begin{eqnarray}
\mathbf{G}_{\mathrm{Zeno}} \sim \left( 
\hat{S}_{\mathtt{zz}},\hat{L}_{\mathtt{z}},\hat{H}\right) .
\end{eqnarray}

\begin{theorem}[Bouten and Silberfarb 2008 \protect\cite{Bouten_Silberfarb}]

Suppose we are given a sequence of bounded operator parameters $\left( S,L\left( k\right) ,H\left( k\right) \right) $ satisfying the scaling,
kernel and decoupling conditions. Then the quantum stochastic process $U_{k}\left( t\right) P_{\mathtt{z}}$ converges strongly to $\hat{U}\left(
t\right) P_{\mathtt{z}}$, that is 
\begin{equation*}
\lim_{k\rightarrow \infty }\left\| U_{k}\left( t\right) \psi -U\left(
t\right) \psi \right\| =0
\end{equation*}
for all $\psi \in \mathfrak{h}\otimes \mathfrak{F}$ with $P_{\mathtt{f}}\otimes I_{\mathfrak{F}}\,\psi =0$. The convergence is moreover uniform in the
time coordinate $t$ for compact time sets.
\end{theorem}

The restriction to bounded operators was lifted in a subsequent publication 
\cite{Bouten_Handel_Silberfarb}.

\begin{definition}
Let $\left( S(k),L(k),H(k)\right) $ be a sequence of $SLH$ operators based on a given space $\mathfrak{h}$ having the dependence on a scaling parameter 
$k>0$ as in equation (\ref{eq:scaling1}). A subspace $\mathfrak{h}_{\mathrm{Zeno}}$ is said to be \textbf{zenofiable} if, with respect to the orthogonal
decomposition $\mathfrak{h}=\mathfrak{h}_{\mathrm{Zeno}}\oplus \mathfrak{h}_{\mathrm{Fast}}$, the scaling, kernel and decoupling conditions are
satisfied. The subspace $\mathfrak{h}_{\mathrm{Zeno}}$ is called the \textbf{Zeno subspace} for the $k\uparrow \infty $ asymptotic dynamics.
The operators $\left( 
\hat{S}_{\mathtt{zz}},\hat{L}_{\mathtt{z}},\hat{H}\right) $ are called the \textbf{Zeno scattering matrix}, the \textbf{Zeno coupling (or collapse) operators} and the
\textbf{Zeno Hamiltonian} operators respectively.
\end{definition}

\section{Controllability and Observability Issues}
The issue of controllability and the Zeno effect has been recently addressed by Burgarth \textit{et al.} \cite{Plato} who have shown that for a system with controlled Hamiltonian of the form
$ \sum_{k=1}^n H_k \, u_k (t)$, with deterministic control policies $u_k$ and self-adjoint operators $H_k$, there are some unexpected benefits of constraints. In particular,
for the case dim $\mathfrak{h} < \infty$, they show that the degree of controllability of the original system may be smaller than the degree of controllability of the Zeno
limit. That is, the dimension of the Lie algebra $\mathfrak{L}_{\mathtt{z}}$ generated by 
$\{ - i P_{\mathtt{z}}H_1 P_{\mathtt{z}} , \cdots , -i P_{\mathtt{z}} H_n P_{\mathtt{z}} \}$ may be strictly greater than that of $\mathfrak{L}$ generated by $\{ - iH_1 , \cdots , -iH_n \}$,
despite the fact that the Zeno dynamics is constrained to the smaller subspace.

We should also remark that an experiment to obtain a limiting Zeno dynamical behaviour for a cavity mode through interaction with and repeated measurements of Rydberg atom 
has been proposed by the Ecole Normale Sup\'{e}rieure group \cite{Zeno_Cavity}.

\subsection{Zeno Master Equation}

Let us prepare the system in an initial state $\eta \in \mathfrak{h}_{\text{Zeno}}$, then for $t\geq 0$ we have the expectation of an observable $X$ on the Zeno subspace given by
\begin{equation*}
\mathbb{E} _{t}( X) = \langle \eta \otimes \Omega | j_t (X) \, \eta \otimes \Omega \rangle  \equiv tr_{\mathtt{z}}\left\{ \rho _{\mathtt{z}}X\right\} ,
\end{equation*}
which introduces the Zeno density matrix $\rho _{\mathtt{z}}$. (The trace is over the Zeno subspace $\mathfrak{h}_{\mathrm{Zeno}}$.) We obtain the Zeno-Ehrenfest equation 
\begin{equation*}
\frac{d}{dt}\mathbb{E} _{t} ( X) =\mathbb{E} _{t}^{\text{vac}}( \mathscr{L}_{\text{Zeno}}X) ,
\end{equation*}
where the Zeno Lindbladian is
\begin{equation*}
 \mathscr{L}_{\text{Zeno}}X =  \hat{L}^\ast_{\mathtt{z} } X \hat{L}_{\mathtt{z}}  -\frac{1}{2} X \hat{L}^\ast_{\mathtt{z}} \hat{L}_{\mathtt{z}}
-\frac{1}{2}  \hat{L}^\ast_{\mathtt{z} } \hat{L}_{\mathtt{z} } X
-i\left[ X ,\hat{H} \right] .
\end{equation*}
The equivalent master equation is
\begin{equation}
\frac{d}{dt}\rho _{\mathtt{z}} = \mathscr{D}_{\mathrm{Zeno}} \rho _{\mathtt{z}}  ,
\end{equation}
where $\mathscr{D}_{\mathrm{Zeno}} \rho _{\mathtt{z}} 
=\hat{L}_\mathtt{z} \rho_\mathtt{z} \hat{L}^\ast_\mathtt{z}-\frac{1}{2} \rho_\mathtt{z} \hat{L}^\ast_\mathtt{z} \hat{L}_\mathtt{z}
-\frac{1}{2}  \hat{L}^\ast_\mathtt{z} \hat{L}_\mathtt{z}  \rho_\mathtt{z}
+i\left[ \rho_\mathtt{z} ,\hat{H} \right] $.

\subsection{Zeno Quantum Trajectories for Vacuum Input}

We now consider possible continual measurements of the output field. We set 
\begin{enumerate}
\item  Homodyne $Z( t) =B_{\text{in}}( t) +B_{\text{in}}^{\ast }( t) ;$

\item  Number counting $Z( t) = \Lambda_{\text{in}}( t) .$
\end{enumerate}

In both cases, the family $\left\{ Z( t) :t\geq 0\right\} $ is self-commuting, as is the set of observables 
\begin{equation}
Y( t) =V( t) ^{\ast }( I_{\text{sys}}\otimes
Z( t) ) V( t) 
\end{equation}
which constitute the actual measured process. We note the non-demolition property $\left[ j_{t}(X),Y_{s}\right] =0$ for all $t\leq s$.

\begin{figure}[htbp]
	\centering
		\includegraphics[width=1.00\textwidth]{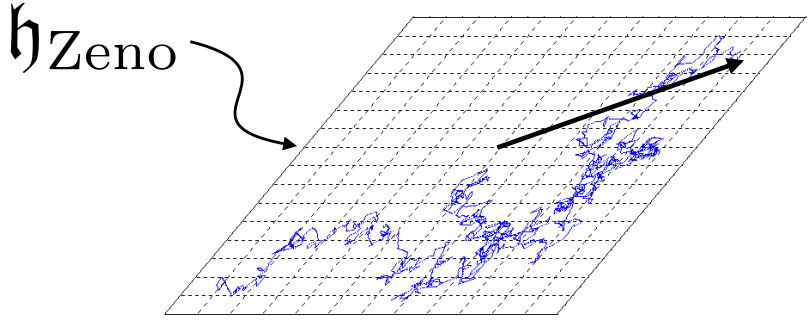}
	\caption{We may still measure the environment and obtain a conditioned state for the system within the Zeno subspace $\mathfrak{h}_{\mathrm{Zeno}}$. }
	\label{fig:Zeno1}
\end{figure}

The aim of filtering theory is to obtain a tractable expression for the least-squares estimate of $j_{t}( X) $ given the output observations $Y( \cdot ) $ up to time $t$. Mathematically, this
is the conditional expectation
\begin{equation*}
\pi _{t}( X) =\mathbb{E}_{\eta \Omega }\left[ j_{t}(
X) |\mathcal{Y}_{t}\right] 
\end{equation*}
onto the measurement algebra $\mathcal{Y}_{t}$ generated by $Y( s) $ for $s\leq t$, with the fixed state $\eta \otimes \Omega$.

The Zeno filters are given respectively by 
\begin{equation}
d\hat{\pi}_{t}(X)=\hat{\pi}_{t}(\mathcal{L}_{\text{Zeno}}X)dt+\mathcal{\hat{H}}_{t}\left( X\right) d\hat{I}(t)
\end{equation}
where we have
\begin{equation*}
\begin{tabular}{l|l|l|}
& Homodyne & Number Counting \\ \hline
$\mathcal{\hat{H}}_{t}\left( X\right) $ & $\hat{\pi}_{t}(X\hat{L}+\hat{L}^{\ast }X)-\hat{\pi}_{t}(X)\hat{\pi}_{t}(\hat{L}+\hat{L}^{\ast })$ & 
$dY(t)-\hat{\pi}_{t}(\hat{L}+\hat{L}^{\ast })dt$ \\ \hline
$d\hat{I}(t)$ & $\hat{\pi}_{t}(\hat{L}^{\ast }X\hat{L})/\hat{\pi}_{t}(\hat{L}^{\ast }\hat{L})-\hat{\pi}_{t}(X)$ & $dY(t)-\hat{\pi}_{t}(\hat{L}^{\ast }\hat{L})dt$ \\ \hline
\end{tabular} .
\end{equation*}

The stochastic processes $\hat{I}$ are the innovations and correspond to the difference between observations $dY$ and the expectations given the present conditioned state.

Alternatively we may introduce the conditioned Zeno state $\hat{\varrho} (t)$ defined by $\hat{\pi}_t (X)  \equiv tr_{\mathtt{z}}\left\{ \hat{ \varrho }_{\mathtt{z}}X\right\}$
for any bounded operator $X$ on the Zeno subspace.
From the filter equation we deduce the Zeno stochastic master equations
\begin{eqnarray}
\hat{\varrho} (t) = \mathscr{D}_{\mathrm{Zeno} }\hat{\varrho} (t) \, dt
+ \hat{\mathcal{G} } ( \hat{\varrho} (t)) \, d\hat{I} (t),
\end{eqnarray}
where 
\begin{equation*}
\begin{tabular}{l|l|l|}
& Homodyne & Number Counting \\ \hline
$\mathcal{\hat{G}}_{t}\left( \varrho \right) $ & $ \hat{L} \varrho + \varrho \hat{L}^{\ast }-tr_{\mathtt{z}}\left\{   \varrho (\hat{L} +\hat{L}^\ast ) \right\} 
\varrho$ & 
$dY(t)-  tr_{\mathtt{z}}\left\{   \varrho  (\hat{L}+\hat{L}^{\ast }) ) \right\} \, dt$ \\ \hline
$d\hat{I}(t)$ & $ \hat{L} \varrho \hat{L}^{\ast } / tr_{\mathtt{z}}\left\{ \hat{ \varrho } \hat{L}^\ast \hat{L} \right\} -\varrho $ & 
$dY(t)-tr_{\mathtt{z}}\left\{   \varrho   \hat{L}^\ast \hat{L} \right\}  dt$ \\ \hline
\end{tabular} .
\end{equation*}

\subsection{Zeno Dynamics Within Networks}
Let $\mathbf{G}_\alpha (k)$ be a collection of zenofiable models with state spaces $\mathfrak{h}_\alpha$,
Zeno spaces $\mathfrak{h}_{\mathrm{Zeno}, \alpha}$, and limit Zeno generators $\mathbf{G}_{\mathrm{Zeno}, \alpha}$.

It is possible to form a quantum feedback network \cite{GJ09}, \cite{GoughJamesIEEE09} by connecting the output fields of
some of these components as the inputs to others. In the limit of instantaneous feedback/forward we obtain a single
effective $SLH$ model for the remaining inputs. The resulting model will again be zenofiable with network Zeno subspace
\[
\mathfrak{h}_{\mathrm{Zeno}} = \bigotimes_\alpha \mathfrak{h}_{\mathrm{Zeno}, \alpha}.
\]
The consistency of the construction follows from earlier work which establishes that the feedback reduction procedure for
determining network models commutes with the procedure for adiabatic elimination \cite{GNW}, \cite{GN}. We obtain the same Zeno dynamics by first 
reducing the components systems to their Zeno dynamics and then performing the network interconnections.

\section{Examples}
In this section we discuss some well-known examples from the perspective of Zeno dynamics and control theory.
\subsection{No scattering, and trivial damping}
Let us set $S=I$, $L^{(1)}=0$, $L^{(0)}_{\mathtt{fz}}=0$ and $H^{(1)}_{\mathtt{zf}} =H^{(1)\ast}_{\mathtt{fz}}=0$. In this case the only 
damping of significance is that of the Zeno component.
Then we have
$A_{\mathtt{ff}} = -i H^{(2)}_{\mathtt{ff}}$ and we require that $H^{(2)}_{\mathtt{ff}}$ is invertible on $\mathfrak{h}_{\mathtt{z}}$.
It is easy to see that the decoupling conditions now apply and we obtain the open Zeno dynamics with $(\hat{S}=I, \hat{L}= L^{(0)}_{\mathtt{zz}}, \hat{H})$
where the Zeno Hamiltonian is
\[
\hat{H}=H_{\mathtt{zz}}^{\left( 0\right) },
\]
and we note that 
\[
\left[ 
\begin{array}{ll}
\hat{H} & 0 \\ 
0 & 0
\end{array}
\right] \equiv P_{\mathtt{z}} H(k) P_{\mathtt{z}} .
\]

The conditions are however still met if we take the off-diagonal terms $H^{(1)}_{\mathtt{zf}} =H^{(1)\ast}_{\mathtt{fz}}$ to be non-zero, and here we have
the Zeno Hamiltonian
\begin{equation*}
\hat{H}=H_{\mathtt{zz}}^{\left( 0\right) }-H_{\mathtt{zf}
}^{(1)}\frac{1}{H_{\mathtt{ff}}^{(2)}}H_{\mathtt{fz}}^{\left( 1\right) }
\end{equation*}
Now $\hat H$ is the shorted version (Schur complement) of 
$H(1)=\left[ 
\begin{array}{cc}
H_{\mathtt{zz}}^{\left( 0\right) } & H_{\mathtt{zf}}^{(1)} \\ 
H_{\mathtt{fz}}^{\left( 1\right) } & H_{\mathtt{ff}}^{(2)}
\end{array}
\right] $.
Equivalently, $\hat H$ is the the limit $k \uparrow \infty$ of shorted version of $H(k)$.

\subsubsection{Qubit Limit}

Let us consider a cavity consisting of a single photon mode with annihilator 
$a$, so that $\left[ a,a^{\ast }\right] =I$. The number states $|n\rangle $, $\left( n=0,1,\cdots \right) $, span the infinite dimensional Hilbert space.
The following model due to Mabuchi \cite{Mabuchi_2012} shows how a large Kerr non-linearity
leads to a Zeno dynamics where we are restricted to the ground and first
excited state of the mode, and so have an effective qubit dynamics. We
consider the $n=2$ input model with
\begin{eqnarray*}
\left[ S(k)\right] _{jk} &=&\delta _{jk}I, \\
\left[ L\left( k\right) \right] _{j} &=&\sqrt{\kappa _{j}}e^{i\omega
t}a,\quad (j=1,2) \\
H\left( k\right)  &=&k^{2}\chi _{0}a^{\ast 2}a^{2}+\Delta a^{\ast }a -i \sqrt{\kappa _{1}} \left( \alpha \left( t\right) a^{\ast }-\alpha ^{\ast
}\left( t\right) a\right) .
\end{eqnarray*}
In the model we are in a rotating frame with frequency $\omega $ and the cavity is detuned from this frequency by an amount $\Delta $. There is a
Kerr non-linearity of strength $\chi \left( k\right) =\chi _{0}k^{2}$ which will be the large parameter. We have two input fields with damping rate 
$\kappa _{j} \, (j=1,2)$, and the first input introduces a coherent driving field $\alpha \left( t\right) $. 

We have $A\equiv \chi _{0}a^{\ast 2}a^{2}=\chi _{0}N\left( N-1\right) $ where $N=a^{\ast }a$ is the number operator. The kernel space of $A$ is
therefore
\begin{equation*}
\mathfrak{h}_{\mathrm{\mathtt{z}}}=\text{span }\left\{ |0\rangle ,|1\rangle
\right\} .
\end{equation*}
For this situation we have $P_{\mathtt{z}}=|0\rangle \langle 0|+|1\rangle \langle 1|$, and we find $L_{\mathtt{fz}}^{(0)}=0$ since 
$P_{\mathtt{f}}aP_{\mathtt{z}}\equiv 0$. The zenofiability conditions are then satisfied and we have
\begin{equation*}
H_{\mathtt{zz}}^{(0)}=P_{\mathtt{z}}\Delta a^{\ast }aP_{\mathtt{z}}\equiv
\Delta \sigma ^{\ast }\sigma 
\end{equation*}
where $\sigma \triangleq P_{\mathtt{z}}aP_{\mathtt{z}}\equiv |0\rangle
\langle 1|$. We then have that
\begin{eqnarray*}
\left[ \hat{S}_{\mathtt{zz}}\right] _{jk} &=&\delta _{jk}\,I_{\mathtt{z}}, \\
\left[ \hat{L}_{\mathtt{z}}\right] _{j} &=&\sqrt{\kappa _{j}}e^{i\omega
t}\sigma , \\
\hat{H} &=&\Delta \sigma ^{\ast }\sigma -i \sqrt{\kappa _{1}} \left(
\alpha \left( t\right) \sigma ^{\ast }-\alpha ^{\ast }\left( t\right) \sigma
\right) .
\end{eqnarray*}
The system is then completely controllable through the policy $\alpha$, and observable through
quadrature measurement (homodyning with $B_{\mathrm{out},1} (t) -B_{\mathrm{out},1} (t)^\ast$, and $-i B_{\mathrm{out} ,1} (t) +i B_{\mathrm{out},1} (t)^\ast$)
and by photon counting.

\subsection{No scattering, but non-trivial damping}

We consider the case where $S=I$,  $L_{\mathtt{fz}}^{(0)}=0$ and  $L_{\mathtt{ff}}^{(1)}=0$, but  $L_{\mathtt{zf}}^{(1)}\neq 0$. The decoupling
conditions are automatically satisfied, so this set-up is zenofiable with Zeno provided that $A_{\mathtt{ff}}$, which is now given by
\begin{equation*}
A_{\mathtt{ff}}\equiv -\frac{1}{2}L_{\mathtt{zf}}^{\left( 1\right) \ast }L_{\mathtt{zf}}^{\left( 1\right) }-iH_{\mathtt{ff}}^{\left( 2\right) },
\end{equation*}
is invertible. If so the Zeno $SLH$ takes the simplified form
\begin{eqnarray*}
\hat{S}_{\mathtt{zz}} &\equiv &I_{\mathtt{z}}+L_{\mathtt{zf}}^{\left(
1\right) }\frac{1}{A_{\mathtt{ff}}}L_{\mathtt{zf}}^{\left( 1\right) \ast },
\\
\hat{L}_{\mathtt{z}} &\equiv &L_{\mathtt{zz}}^{\left( 0\right) }-L_{\mathtt{zf}}^{\left( 1\right) }\frac{1}{A_{\mathtt{ff}}}M_{\mathtt{fz}}, \\
\hat{H} &\equiv &H_{\mathtt{zz}}^{(0)}+\mathrm{Im}\left\{ M_{\mathtt{zf}} \frac{1}{A_{\mathtt{ff}}}M_{\mathtt{fz}}\right\} ,
\end{eqnarray*}
where now 
\begin{eqnarray*}
M_{\mathtt{zf}} &\equiv &-\frac{1}{2}L_{\mathtt{zz}}^{\left( 0\right) \ast
}L_{\mathtt{zf}}^{\left( 1\right) }-iH_{\mathtt{zf}}^{\left( 1\right) }, \\
M_{\mathtt{fz}} &\equiv &-\frac{1}{2}L_{\mathtt{zf}}^{\left( 1\right) \ast
}L_{\mathtt{zz}}^{\left( 0\right) }-iH_{\mathtt{fz}}^{\left( 1\right) }.
\end{eqnarray*}

\subsubsection{Alkali Atom}

Consider a model for an atomic electron which has two energy states, a ground state $|g\rangle $ and an excited state $|e\rangle $ spanning 
$\mathfrak{h}_{\text{level}}=\mathbb{C}^{2}$, and an intrinsic spin one-half with states 
$|+\rangle $ and $|-\rangle $ spanning $\mathfrak{h}_{\text{spin}}=\mathbb{C}^{2}$. 
The total Hilbert space is then the 4-dimensional $\mathfrak{h}=\mathfrak{h}_{\text{level}}\otimes \mathfrak{h}_{\text{spin}}$. Taking $n=3$ inputs, one for
each spatial coordinate, the scaled model is \cite{BHJ07}
\begin{eqnarray*}
\left[ L(k)\right] _{j} &=&k\sqrt{\gamma }|g\rangle \langle e|\otimes \sigma
_{j}\text{, \ \ }j=1,2,3, \\
H\left( k\right)  &=&k^{2}\Delta |e\rangle \langle e|\otimes
I+\sum_{j=1,2,3}I\otimes \mathcal{B}_{j}\sigma _{j}
\end{eqnarray*}
where $\Delta >0$ has the interpretation of a detuning frequency for the
excited state, and $\mathcal{B}_{j}$ as the $j$ component of an external
magnetic field. We find that
\begin{equation*}
A=-\left( \frac{3}{2}\gamma +i\Delta \right) |e\rangle \langle e|\otimes I
\end{equation*}
so if we set $P=|g\rangle \langle g|\otimes I$, we have a zenofiable model
with 2-dimensional Zeno space
\begin{equation*}
\mathfrak{h}_{\text{Zeno}}=|g\rangle \otimes \mathfrak{h}_{\text{spin}}\equiv \text{%
span}\left\{ |g\rangle \otimes |+\rangle ,|g\rangle \otimes |-\rangle
\right\} .
\end{equation*}
The Zeno $SLH$ is
\begin{eqnarray*}
\left[ \hat{S}\right] _{jk} &=&|g\rangle \langle g|\otimes \left\{ \delta
_{jk}-\frac{\gamma }{\frac{3}{2}\gamma +i\Delta }\sigma _{j}\sigma
_{k}\right\} , \\
\left[ \hat{L}\right] _{j} &=&0, \\
\hat{H} &=&\sum_{j=1,2,3}|g\rangle \langle g|\otimes \mathcal{B}_{j}\sigma
_{j}.
\end{eqnarray*}

In this case the Zeno dynamics is undamped $(\hat{L}=0)$ so filtering is not possible for vacuum input. However the Zeno system (effectively the spin) is
controllable through the magnetic field. 

\subsubsection{$\Lambda $-systems}

Consider a three level atom with ground states $|g1\rangle $,$|g2\rangle $
and an excited state $|e\rangle $ with Hilbert space $\mathfrak{h}_{\text{level}}=\mathbb{C}^{3}$. The atom is contained in a cavity with quantum mode $a$
with Hilbert space $\mathfrak{h}_{\text{mode}}$ where $\left[ a,a^{\ast }\right]
=1$ and $a$ annihilates a photon of the cavity mode. The combined system and
cavity has Hilbert space $\mathfrak{h}=\mathfrak{h}_{\text{level}}\otimes \mathfrak{h}_{\text{mode}}$, and consider the following \cite{DuanKimble}, \cite{BHJ07},
\begin{eqnarray*}
L\left( k\right)  &=&k\sqrt{\gamma }I\otimes a, \\
H(k) &=&ik^{2}\mathsf{g}\left\{ |e\rangle \langle g1|\otimes a-|g1\rangle
\langle e|\otimes a^{\ast }\right\} +ik\left\{ |e\rangle \langle g2|\otimes
\alpha -|g2\rangle \langle e|\otimes \alpha ^{\ast }\right\} .
\end{eqnarray*}
Here the cavity is lossy and leaks photons with decay rate $\gamma $, we
also have a transition from $|e\rangle $ to $|g1\rangle $ with the emission
of a photon into the cavity, and a scalar field $\alpha $ driving the
transition from $|e\rangle $ to $|g2\rangle $. We see that
\begin{equation*}
A\equiv -\frac{1}{2}\gamma I\otimes a^{\ast }a+\mathsf{g}\left\{ |e\rangle
\langle g1|\otimes a-|g1\rangle \langle e|\otimes a^{\ast }\right\} .
\end{equation*}
and that $A$ has a 2-dimensional kernel space spanned by the pair of states
\begin{equation*}
|\Psi _{1}\rangle =|g1\rangle \otimes |0\rangle ,\quad |\Psi _{2}\rangle
=|g2\rangle \otimes |0\rangle .
\end{equation*}
The Zeno subspace is then the span of $|\Psi _{1}\rangle $ and $|\Psi
_{2}\rangle $, and the resulting $SLH$ operators are
\begin{eqnarray*}
\hat{S} &=&|\Psi _{1}\rangle \langle \Psi _{1}|-|\Psi _{2}\rangle \langle
\Psi _{2}|  \equiv I-2 \sigma^\ast \sigma , \\
\hat{L} &=&-\frac{\gamma \alpha }{\mathsf{g}}|\Psi _{1}\rangle \langle \Psi _{2}|  \equiv -\frac{\gamma \alpha }{\mathsf{g}}  \sigma , \\
\hat{H} &=&0 ,
\end{eqnarray*}
where $\sigma = |\Psi _{1}\rangle \langle \Psi _{2}| $.
Here the Zeno dynamics has a vanishing Zeno Hamiltonian, but is partially
observable through filtering as $\hat{L}\neq 0$.

\section{Oscillator Models}
The original treatment of adiabatic elimination problem for quantum stochastic problems dealt with
systems coupled to oscillator components which decayed rapidly to their ground states \cite{GvH}. 
Decomposing the system Hilbert space as $\mathfrak{h} = \mathfrak{h}_{\mathrm{slow}} \otimes \mathfrak{h}_{\mathrm{osc}}$,
one can establish the conditions for adiabatic elimination of the oscillators lead to the Zeno subspace
\begin{equation}
\mathfrak{h}_{\mathrm{Zeno}} = \mathfrak{h}_{\mathrm{slow}} \otimes \vert 0 \rangle ,
\label{eq:osc_zeno}
\end{equation}
where $\vert 0 \rangle $ is the ground state of the oscillatory system. The effective Zeno
subspace obviously being the slow space. 

We consider an open model described by the generators $\mathbf{G} (k)$ with
\begin{eqnarray}
S(k) &=& \hat S \otimes I,  \notag \\
L(k) &=&k\sum_{i} \hat C_{i}\otimes a_{i}+ \hat G\otimes I,  \notag \\
K(k) &=&k^{2}\sum_{ij}\hat A_{ij}\otimes a_{i}^{\ast }a_{j}+k\sum_{i}\hat Z_{i} \otimes
a_{i}^{\ast }+k\sum_{i} \hat X_{i}\otimes a_{i}+ \hat R \otimes I,
\label{scaled parameters}
\end{eqnarray}
where $k$ is a positive scaling parameter and $\hat S , \hat C_{i}, \hat G , \hat A_{ij} , \hat X_{i} , \hat Z_{i} , \hat R
$ are bounded operators on $\mathfrak{h}_{\mathrm{slow}}$ with $\hat A=\left[ \hat A_{ij}\right]
$ invertible with bounded inverse. Here $a_{i}$ is the annihilator corresponding to the 
$i$th local oscillator, say with $i=1,\cdots ,m$.

As $k\rightarrow \infty $ the oscillators become increasingly strongly coupled to the driving noise field and in this limit we would like to
consider them as being permanently relaxed to their joint ground state. The oscillators are then the fast degrees of freedom of the system, with the
auxiliary space $\mathfrak{h}_{\mathrm{slow}}$ describing the slow degrees. 

We say that $\hat A$ is \textit{strictly Hurwitz stable} if 
\begin{equation*}
\mathrm {Re}\langle \psi |\hat A \psi \rangle <0\text{, \ for all }\psi \neq 0.
\label{Hurwitz}
\end{equation*}

It is shown in \cite{GNW} that $A = \sum_{ij}\hat A_{ij}\otimes a_{i}^{\ast }a_{j}$ will have kernel space 
given by (\ref{eq:osc_zeno}) whenever 
$\hat A$ is strictly Hurwitz. Assuming that $\hat A$ is Hurwitz, we then obtain a Zeno limit with $\mathbf{G}_{\mathrm{Zeno}}$
\begin{eqnarray}
\hat{S} &=&(I+\hat C{\hat A}^{-1} {\hat C}^{\ast }) \hat S,  \notag \\
\hat{L} &=&\hat G- \hat C \hat A^{-1} \hat Z,  \notag \\
\hat{K} &=&\hat R- \hat X \hat A^{-1} \hat Z,  \label{Limit parameters}
\end{eqnarray}
where we drop the ``$\otimes |0 \rangle \langle 0 |_{\mathrm{osc}}$'' for convenience.

\subsection{Quantum Linear Models}
In principle, the slow degrees of freedom may also correspond to
an assembly of oscillators too, and we may take the overall model (fast and slow oscillators) to be
linear. The limit dynamics will then be linear, with the model coefficients determined by the limit theorems. 
These will in fact be similar in structure to singular perturbation results for linear control systems \cite{KKOR}. 
In this case it is known that the strict Hurwitz property of $\hat A$ (now a matrix with scalar entries) 
guarantees that there exist a finite constant $k_0$ such the total linear model is stable for $k >k_0$.
This justifies the neglect of parasitic modes in engineering modelling \cite{Khalil}. For quantum linear passive systems,
this will also be the case, and the non-Zeno part of the dynamics may be safely ignored.

Specifically, we may take the slow system to be comprised of $m$ quantum oscillators with annihilators $b_1 , \cdots , b_r$
and we obtain a linear model with fast oscillators $a_i$ and slow (Zeno!) oscillators $b_i$ by taking
\begin{eqnarray*}
\hat G = \sum_i (G^+_i b_i^\ast +G^-_i \, b_i ),\\
\hat Z_j = \sum_i (Z^+_{ij} b_i^\ast + Z^-_{ij} \, b_i ),\\
\hat X_j = \sum_i (X^+_{ij} b_i^\ast + X^-_{ij} \, b_i ),\\
\hat R_j = \sum_i (R^{++}_{ij} b_i^\ast b_j^\ast  + R^{--}_{ij} \, b_i b_j +R^{+-}_{ij} \, b^\ast_i b_j  ),
\end{eqnarray*}
and the $\hat A_{ij}$ scalars.

From the quantum stochastic calculus \cite{HP} the oscillators satisfy linear Heisenberg equations that take the form
\begin{eqnarray*}
d b (t) = \Gamma_1 b \, dt + \Gamma_2 z + \Phi \, dB_{\mathrm{in}} (t) , \\
\frac{1}{k^2} dz(t) = \Gamma_3 b \, dt + \Gamma_4 z + \Psi \, dB_{\mathrm{in}} (t) ,
\end{eqnarray*}
where we set $z \equiv ka$. The coefficient matrices $\Gamma_j \, (j=1,2,3,4), \Phi , \Psi$ are independent of 
the scaling parameter $k$. In particular $\Gamma_4 \equiv \hat A$.

We are interested primarily in the stability of these equations and it is sufficient to look at the averages of the operators
where the environment state is vacuum, for which the input processes are martingales, and here we have
\begin{eqnarray}
\frac{d}{dt} \bar b (t) = \Gamma_1 \bar b (t) + \Gamma_2 \bar z  (t) , \\ \notag
\frac{1}{k^2} \frac{d}{dt} \bar z(t) = \Gamma_3 \bar b (t) + \Gamma_4 \bar z (t) ,
\label{eq:Gammas}
\end{eqnarray}

\begin{proposition}
If $\hat A$ is invertible, then the necessary and
sufficient conditions for the existence of a $k_0 <\infty$ such that the system (\ref{eq:Gammas}) is asymptotically stable for $k>k_0$ are that 
both $\hat A$ and $\Gamma_0$ are strictly Hurwitz.
\end{proposition}

This is a consequence of the following result \cite{Chow_Kokotovic}.

\begin{lemma} If $\Gamma_4 \equiv \hat A$ is invertible, then for $k$ sufficiently large the eigenvalues for the system of linear dynamical equations (\ref{eq:Gammas})
are $\sigma( \Gamma_0 ) (1 +O( \frac{1}{k^2}) )$, and $ k \sigma( \Gamma_4 ) (1 +O( \frac{1}{k^2}) )$, where $\Gamma_0 = \Gamma_1 - \Gamma_2 \Gamma_4^{-1} \Gamma_3$.
\end{lemma}

(Here $\sigma(X)$ denotes the spectrum of a matrix $X$.) The strict Hurwitz property for $\hat A$ is essential for the existence of the Zeno dynamical limit
for the system of oscillators $b_i$. The additional requirement that $\Gamma_0$ also be Hurwitz is required for stability of the limit system, compare with (\ref{Limit parameters}).

From a modelling perspective, the Zeno dynamics is a model reduction of the total assembly of fast and slow oscillators. In control applications, one would want to
have the performance under the total dynamics approximated arbitrarily closely by the Zeno model by taking $k$ sufficiently large. In this way any further
feedback construction \cite{GJ09} would be robust against the modelling errors introduced by reducing to the Zeno description. To this end we remark again that
feedback in SLH models is compatible with model reduction through adiabatic elimination \cite{GNW,GN}.

\end{document}